\documentclass[12pt]{article}
\usepackage{graphicx}
\def \app{D_{\pi \pi}}

\def \bea{\begin{eqnarray}}
\def \beq{\begin{equation}}

\def \cn{Collaboration}

\def \eea{\end{eqnarray}}
\def \eeq{\end{equation}}
\def \ite{{\it et al.}}

\def \bl{\bar \lambda}

\begin{document}
\begin{flushright}
TECHNION-PH-2004-15\\
CLNS-04-1868 \\
hep-ph/0403287 \\
March 2004 \\
\end{flushright}
\renewcommand{\thesection}{\Roman{section}}
\renewcommand{\thetable}{\Roman{table}}
\centerline{\bf CORRELATED BOUNDS ON CP ASYMMETRIES IN $B^0 \to \eta' K_S$}
\medskip
\vskip3mm
\centerline{Michael Gronau$^a$, Jonathan L. Rosner$^b$\footnote{On leave from
Enrico Fermi Institute and Department of Physics, University of Chicago,
Chicago, Illinois 60637.} and Jure Zupan$^{a, c}$}
\medskip
\vskip3mm
\centerline{$^a$\it Physics Department, Technion -- Israel Institute of Technology}
\centerline{\it 32000 Haifa, Israel}
\medskip
\centerline{$^b$\it Laboratory for Elementary-Particle Physics, Cornell University}
\centerline{\it Ithaca, New York 14850}
\medskip
\centerline{$^c$\it J.~Stefan Institute, Jamova 39, P.O. Box 3000,1001}
\centerline{\it Ljubljana, Slovenia}
\bigskip

\begin{quote}

Flavor SU(3) is used to constrain the coefficients of $\sin\Delta mt$
and $\cos\Delta mt$ in the time-dependent CP asymmetry of $B^0 \to
\eta' K_S$. Correlated bounds in the $(S_{\eta' K},~C_{\eta' K})$
plane are derived, by using recent rate measurements of $B^0$ decays
into $K^+ K^-,~\pi^0\pi^0,~\pi^0\eta,~\pi^0\eta', \eta\eta,
~\eta\eta',~\eta'\eta'$.  Stringent bounds are obtained when assuming a single 
SU(3) singlet amplitude and when neglecting annihilation-type amplitudes.

\end{quote}
\leftline{\qquad PACS codes:  12.15.Hh, 12.15.Ji, 13.25.Hw, 14.40.Nd}
\bigskip

\section{Introduction}
Measurements of time-dependent CP asymmetries in $b \to c \bar cs$
decays including $B^0 \to J/\psi K_S$~\cite{psiKs} are interpreted in the
Standard Model as $\sin 2\beta\sin\Delta mt$, where
$\beta \equiv \arg(-V_{tb}V^*_{td}V_{cd}V_{cb}^*)$. These measurements
have provided a crucial test~\cite{BS} of the Kobayashi-Maskawa
mechanism~\cite{KM}. This test is theoretically clean because a single
weak phase ${\rm arg}(V^*_{cb}V_{cs})$ dominates
$B\to J/\psi K_S$ within a fraction of a percent~\cite{MG,TM}.

An interesting class of processes, susceptible to new physics effects
\cite{NP}, consists of $b \to s$ penguin-dominated $B^0$
decays into CP-eigenstates. This includes the final states $\phi
K_S,~(K^+K^-)_{({\rm even}~\ell)}K_S,~\pi^0 K_S$ and $\eta' K_S$.
Here decay amplitudes contain two terms: a penguin
amplitude, $A'_P$, including a dominant Cabibbo-Kobayashi-Maskawa
(CKM) factor $V^*_{cb}V_{cs}$,  and a color-suppressed tree
amplitude, $A'_C$, with a smaller CKM factor $V^*_{ub}V_{us}$. The first
amplitude by itself would imply a CP asymmetry of magnitude 
$\sin 2 \beta\sin\Delta mt$. The second amplitude modifies 
the coefficient of this term,
and introduces a $\cos\Delta mt$ term in the asymmetry~\cite{MG}. 
The coefficients of
$\sin\Delta mt$ and $\cos\Delta mt$ for a final state $f$ are denoted by
$S_f$ and $-C_f$, respectively. The observables, $\Delta S_f \equiv S_f \pm
\sin 2\beta$ (where the sign depends on the final state CP)
and $C_f$, increase with $|A'_C/A'_P|$, which depends on hadron dynamics,
and are functions of unknown strong interaction phases.
A search for new physics effects in these processes requires a careful
theoretical analysis of $\Delta S_f$ and $C_f$ within the Standard Model.

Model-independent studies of the ratios $|A'_C/A'_P|$ in $B^0\to \phi
K_S$~\cite{GLNQ,CGLRS}, $B^0 \to (K^+K^-)_{({\rm even}~\ell)}$
$K_S$~\cite{GLNQ,GR}, $B^0 \to \pi^0 K_S$~\cite{GGR} and $B^0 \to
\eta'K_S$~\cite{GLNQ,CGR} have been carried out using flavor SU(3).
The idea is simple.  Flavor SU(3) relates the hadronic amplitudes $A'_P$ and 
$A'_C$ in each of these processes to corresponding classes of hadronic
amplitudes in $\Delta S = 0$ decay processes. Strangeness-changing and
strangeness-conserving amplitudes include CKM factors which satisfy
well-defined ratios. Consequently, rate measurements in the $\Delta S =0$
sector may provide bounds on ratios of amplitudes $|A'_C/A'_P|$ in $\Delta S
= 1$ decays.

In the present Letter we will calculate correlated bounds on 
$\Delta S_{\eta' K}$ and $C_{\eta' K}$ in $B^0(t) \to \eta'K_S$,
using very recent branching ratio measurements of $B^0$ decays into
a pair of neutral charmless light pseudoscalars. The two 
asymmetries are proportional to the ratio $|A'_C/A'_P|$ in this process. 
Approximate bounds on $|A'_C/A'_P|$, based on flavor SU(3), were 
presented in \cite{GLNQ} and in the appendix of \cite{CGR} using earlier 
data. These bounds neglected interference effects and ${\cal O}(\lambda^2)$
terms ($\lambda = 0.22$) and are expected to hold within a factor of about 1.5. 
The measurables $\Delta S_{\eta'K}$ and $C_{\eta'K}$ 
are of order $|A'_C/A'_P|$. However, they depend also on weak and strong 
phases. An SU(3) method for obtaining correlated bounds on 
$\Delta S_{\pi K}$ and $C_{\pi K}$  in $B^0(t)\to \pi^0 K_S$
was proposed in \cite{GGR},  taking account of this 
dependence and avoiding the above approximation. 
Here we will apply this method to the asymmetries in $B^0(t) \to \eta' K_S$.
This is the first study of correlated bounds on asymmetries in $B^0 \to     %
\eta'K_S$, one of a few processes believed to be penguin-dominated in the   %
Standard Model.                                                             %
 
The bounds obtained here present a great improvement relative to bounds     %
based on earlier data.  This demonstrates the importance of branching ratio %
measurements of processes such as $B \to \eta^{(\prime)}\eta^{(\prime)}$    %
obtained recently.  Current bounds are now at a level where a further       %
improvement in measurements may lead to a deviation from Standard Model     %
predictions.                                                                %

Although the method applied here is based on an idea similar to that        %
applied already in Ref.\ \cite{GGR}, there is one major difference: In $B   %
\to \eta'K_S$ we optimize the bounds using a whole continuum of             %
combinations of processes.  This is a novel feature of the present analysis.%

In Section II we  write expressions for the observed asymmetries in terms of
hadronic amplitudes, noting their dependence on weak and strong phases. 
Section III provides an SU(3) decomposition for the amplitude of $B^0 \to \eta'
K^0$ and for a class of $\Delta S =0$ related processes. In Section IV we
obtain bounds on $\Delta S_{\eta' K}$ and $C_{\eta' K}$ in two ways, both in a
general SU(3) framework and also by neglecting small annihilation-type
amplitudes. Section V concludes by comparing these bounds with expectations in
other approaches, including a recent global SU(3) fit of all $B$ decays to
pairs of charmless pseudoscalars. 

\section{Asymmetries and amplitudes in $B^0 \to \eta'K_S$}
The CP asymmetry in $B^0$ decays to the CP 
eigenstate $\eta'K_S$  has the general expression \cite{MG}:
\beq
A(t) \equiv \frac{\Gamma(\bar B^0(t) \to \eta' K_S) - \Gamma(B^0(t) \to \eta'
K_S)}{\Gamma(\bar B^0(t) \to \eta' K_S) + \Gamma(B^0(t) \to \eta' K_S)}
= -C_{\eta' K} \cos(\Delta mt) + S_{\eta' K}\sin(\Delta mt)~.
\eeq
Measurements obtained by the BaBar \cite{eta'KBaBar} and Belle
\cite{eta'KBelle} Collaborations,
\beq
S_{\eta'K} = \left\{ \begin{array}{c} 0.02 \pm 0.34 \pm 0.03~, \cr
0.43 \pm 0.27 \pm 0.05~,\end{array} \right.
C_{\eta'K} = \left\{ \begin{array}{c} 0.10 \pm 0.22 \pm 0.04~,~~~~
{\rm BaBar}~, \cr
0.01 \pm 0.16 \pm 0.04~,~~~~
{\rm Belle}~,\end{array} \right.
\eeq
imply averages
\beq 
S_{\eta'K} = 0.27 \pm 0.21~,~~~~~~C_{\eta'K} = 0.04 \pm 0.13~.
\eeq

The two measurables $S_{\eta'K}$ and $C_{\eta'K}$ can be expressed in terms 
of the amplitude of $B^0 \to \eta' K^0$.
As mentioned in the introduction, it is convenient to decompose this amplitude
into two terms, $A'_P$ and $A'_C$, involving intrinsic CKM factors 
$V^*_{cb}V_{cs}$ and $V^*_{ub}V_{us}$, and strong and weak
phases $\delta$ and $\gamma$, respectively,
\beq\label{Amp}
A(B^0\to \eta' K^0) = A'_P + A'_C = |A'_P|e^{i\delta} + |A'_C|e^{i\gamma}~.
\eeq
Expressions for $S_{\eta' K}$ and $C_{\eta' K}$ in terms of $A'_P$ and $A'_C$ 
can be obtained from definitions, taking into account the negative 
CP eigenvalue of $\eta' K_S$ in $B^0$ decays \cite{MG}:
\beq
S_{\eta' K} \equiv \frac{2{\rm Im}(\lambda_{\eta' K})}{1 + |\lambda_{\eta' K}|^2}~,
~~~~~~~~~C_{\eta' K} \equiv \frac{1 - |\lambda_{\eta' K}|^2}
{1 + |\lambda_{\eta' K}|^2}~,
\eeq
where
\beq
\lambda_{\eta' K} \equiv -e^{-2i\beta}\frac{A(\bar B^0 \to \eta'
\bar K^0)}{A(B^0 \to \eta' K^0)}~.
\eeq

Using Eq.~(\ref{Amp}), the asymmetries 
$S_{\eta' K}$ and $C_{\eta' K}$ are then written in terms of
$|A'_C/A'_P|,~\delta$, $\gamma$, and $\alpha \equiv \pi - \beta - \gamma$:
\bea
S_{\eta' K} & = & {\sin 2\beta + 2|A'_C/A'_P| \cos \delta \sin(2 \beta +
\gamma) - |A'_C/A'_P|^2 \sin(2 \alpha) \over R_{\eta'K}}~ \label{eqn:S},\\[3pt]
C_{\eta' K} & = & {2|A'_C/A'_P| \sin \delta \sin \gamma \over R_{\eta'K}}~ 
\label{eqn:C},\\[3pt]
R_{\eta'K} & \equiv & 1 + 2|A'_C/A'_P| \cos \delta \cos \gamma +|A'_C/A'_P|^2~.
\label{eqn:R}
\eea
The amplitudes $A'_P$ and $A'_C$ are expected to obey a hierarchy, $|A'_C| \ll
|A'_P|$ \cite{GHLR}.  In the limit of neglecting $A'_C$, one has the well-known
result $S_{\pi K} = \sin 2\beta, C_{\pi K}=0$. Keeping only linear terms in
$|A'_C/A'_P|$, one has \cite{MG}
\bea\label{SC}
\Delta S_{\eta' K} \equiv  S_{\eta' K} - \sin 2\beta & \approx &
2|A'_C/A'_P|\cos 2\beta\cos\delta\sin\gamma~,\nonumber \\
C_{\eta' K} & \approx  & 2|A'_C/A'_P|\sin\delta \sin\gamma~.
\eea
Thus, within this approximation, the allowed region in the $(S_{\eta'K},
C_{\eta'K})$ plane is confined to an ellipse centered at $(\sin 2 \beta, 0)$,
with semi-principal axes $2[|A'_C/A'_P|$ $ \sin \gamma]_{\rm max} \cos 2\beta$
and $2[|A'_C/A'_P| \sin\gamma]_{\rm max}$.  In our study below we will use the
exact expressions (\ref{eqn:S})--(\ref{eqn:R}).

\section{SU(3) decomposition of amplitudes}
A convenient way of introducing flavor symmetry in charmless $B$ decays is
through graphical representations of SU(3) amplitudes~\cite{DZ,Chau,GHLR,DGR}.
This parametrization is equivalent to a pure group-theoretical
presentation~\cite{DZ,SW,Desh,GL}, having the advantage of
anticipating that certain amplitudes are smaller than others.
Our analysis will be carried out both in a general SU(3) framework, and
also neglecting small annihilation-type amplitudes.
We  use quark content for mesons and phase conventions as in \cite{GHLR,DGR}:
\beq
B^0 = d\bar b,~\pi^0=(d\bar d-u\bar u)/\sqrt{2},~K^0=d\bar s,
\eta=(s\bar s-u\bar u-d\bar d)/\sqrt{3},
\eta'=(u\bar u+d\bar d+2s\bar s)/\sqrt{6}~.
\eeq
The $\eta$ and $\eta'$ correspond to octet-singlet mixtures
\beq
\eta  = \eta_8 \cos \theta_0 - \eta_1 \sin \theta_0~,~~
\eta' = \eta_8 \sin \theta_0 + \eta_1 \cos \theta_0~~,
\eeq
with $\theta_0 = \sin^{-1}(1/3) = 19.5^\circ$.

The flavor flow amplitudes, which occur in $B^0$ decays into relevant pairs of
neutral charmless pseudoscalar mesons, are the following: a ``penguin''
contribution $p$; a ``singlet penguin'' contribution $s$, in which a
color-singlet $q \bar q$ pair produced by two or more gluons or by a $Z$ or
$\gamma$ forms an $SU(3)$ singlet state; a ``color-suppressed'' contribution
$c$;  an ``exchange'' contribution $e$,  and a ``penguin annihilation''
contribution $pa$. The three amplitudes, $p,~s$ and $c$ contain both 
leading-order and electroweak penguin contributions \cite{GHLR}. 
We shall denote $\Delta S = 0$ transitions by unprimed quantities, $p, s, c, e$
and $pa$, and $|\Delta S| = 1$ transitions by corresponding primed quantities.

We note that in a general SU(3) analysis decays of $B$ mesons into pairs of
pseudoscalars, consisting of a singlet and an octet of SU(3), are described by
three SU(3) amplitudes~\cite{DZ}. Our parametrization uses the single
amplitude $s$, neglecting two other amplitudes in which the spectator quark
enters the decay Hamiltonian~\cite{DGR}.  The other amplitudes, in which the
spectator quark participates in decay processes, are $e$ and $pa$.  These
amplitudes may be assumed to be smaller than the others \cite{GHLR}, and will
be neglected in part of our discussion.  Experimental evidence for the
suppression of the combination $e + pa$, already exhibited by the current upper
bound on $B^0 \to K^+K^-$ \cite{PDG}, is expected to be strengthened in future
measurements.  We expect the approximation involved in neglecting small
amplitudes to be comparable to that associated with assuming flavor SU(3)
symmetry.  For generality, we will also give exact results within SU(3)
which do not neglect small contributions.

Expressions for decay amplitudes in terms of graphical contributions are
obtained in a straightforward manner~\cite{GHLR,DGR}. For  $B^0 \to \eta'K^0$ 
one finds
\beq\label{eta'K}
\sqrt{6}A(B^0 \to K^0\eta') = 3p' + 4s' + c' \equiv A'_P + A'_C.
\eeq
Similarly, one finds expressions for a set of SU(3) related
strangeness-conserving amplitudes of which we list those 
that are useful for constraining the asymmetry in 
$B^0 \to \eta'K_S$:
\bea\label{pieta}
A(B^0 \to K^+K^-) & = & -e -pa~,\nonumber\\
A(B^0 \to K^0\bar K^0) & = & p + pa~,\nonumber\\
\sqrt{2}A(B^0 \to \pi^0\pi^0) & = & p - c + e + pa~,\nonumber\\
\sqrt{6}A(B^0 \to \pi^0\eta) & = & - 2p - s + 2e~,\nonumber\\
\sqrt{3}A(B^0 \to \pi^0\eta') & = & p + 2s - e~,\nonumber\\
(3/\sqrt{2})A(B^0 \to\eta\eta) & = & p + s + c + e + (3/2)pa~,\nonumber\\
3\sqrt{2}A(B^0\to \eta'\eta') & = & p + 4s + c + e + 3pa~,\nonumber\\
3\sqrt{2}A(B^0\to \eta\eta') & = & -2p - 5s - 2c - 2e~.
\eea
We denote the amplitudes of these processes by $A(f)$, where $f$
stands for a given final state.
Eqs.~(\ref{eta'K})--(\ref{pieta}) provide a starting point for our analysis,
in which correlated bounds on $\Delta S_{\eta' K}$ and $C_{\eta' K}$ will be
obtained in terms of rate measurements of the processes occurring in
(\ref{pieta}).

\section{Correlated bounds on $\Delta S_{\eta' K}$ and $C_{\eta' K}$}
The basis of potential bounds on $\Delta S_{\eta' K}$ and $C_{\eta' K}$
is the identical SU(3) structure of the amplitude of $B^0\to \eta'K^0$
(\ref{eta'K}) and that of certain linear combinations of the amplitudes
(\ref{pieta}).  The role of such a relation in setting approximate bounds on
$|A'_C/A'_P|$ was pointed out in \cite{GLNQ}. The method for deriving
correlated bounds on the asymmetries $S$ and $C$ has already been applied to
$B^0\to \pi^0 K_S$ \cite{GGR}. In order to implement the method in $B^0 \to
\eta'K_S$, one is searching for a linear superposition of the $\Delta S =0$
amplitudes $A(f)$ in (\ref{pieta}), with given real coefficients $a_f$,
which acquires an expression similar to (\ref{eta'K}),
\beq\label{psc}
\sqrt{6}\Sigma_f a_fA(f) = 3p + 4s + c \equiv A_P + A_C~.
\eeq
The amplitudes $A_P$ and $A_C$ include CKM factors $V^*_{cb}V_{cd}$ 
and $V^*_{ub}V_{ud}$, respectively.
The flavor SU(3) structures of (\ref{eta'K}) and (\ref{psc}) are identical,
while their CKM factors satisfy simple well-defined ratios,
\bea\label{eqn:CKM}
\frac{A_P}{A'_P} & = & \frac{V^*_{cb}V_{cd}}{V^*_{cb}V_{cs}} 
= -\bl~,\nonumber\\
\frac{A_C}{A'_C} & = & \frac{V^*_{ub}V_{ud}}{V^*_{ub}V_{us}} 
= \bl^{-1}~,\nonumber\\
\bl & = & \frac{\lambda}{1- \lambda^2/2} = 0.230~.
\eea

Since the eight physical amplitudes $A(f)$ are expressed in terms of five SU(3)
contributions, one may form a whole continuum of combinations satisfying 
(\ref{psc}). Constraints
on the asymmetry in $B^0 \to \eta'K_S$ will be shown to follow from upper
bounds on rates of processes appearing on the left-hand-side of (\ref{psc}). 
The choice of a combination leading to the strongest constraints on the
asymmetry depends on experimental upper bounds available at a given time.
Three cases are of particular interest because of their current implications on
the $\eta'K$ asymmetry:
\begin{enumerate}
\item One combination, involving pairs including $\pi^0,~\eta$ and $\eta'$ in
the final state, was proposed in~\cite{GLNQ} by using a complete SU(3)
analysis, and in~\cite{CGR} by applying simple U-spin symmetry arguments:
\bea\label{six}
\Sigma_f a_fA(f)& = &
\frac{1}{4\sqrt{3}}A(\pi^0\pi^0) - \frac{1}{3}A(\pi^0\eta) +
\frac{5}{6\sqrt{2}}A(\pi^0\eta')
\nonumber\\
& + & \frac{2}{3\sqrt{3}}A(\eta\eta)  - \frac{11}{12\sqrt{3}}A(\eta'\eta') -
\frac{5}{3\sqrt{3}}A(\eta\eta')~.
\eea
\item Another combination, based on the assumption that a single 
SU(3) amplitude dominates decays into a singlet and an octet 
pseudoscalar, involves four decay processes including $B^0 \to K^+ K^-$: 
\beq\label{four}
\Sigma_f a_fA(f)  =  \frac{1}{3\sqrt{3}}A(\pi^0\pi^0) + \frac{1}{3\sqrt{6}}
A(K^+K^-) - \frac{2}{3}A(\pi^0\eta) - \frac{2}{\sqrt{3}}A(\eta\eta')
\eeq
\item A third superposition, satisfying (\ref{psc}) in the limit $e = pa = 0$,
involves only three strangeness-conserving amplitudes:
\beq\label{three}
\Sigma_f a_fA(f) = -\frac{5}{6}A(\pi^0\eta) + \frac{1}{3\sqrt{2}}A(\pi^0\eta')
- \frac{\sqrt{3}}{2}A(\eta\eta')~.
\eeq
\end{enumerate}
The coefficients $a_f$ in these three cases can be read off Eqs.~(\ref{six}),
(\ref{four}), and (\ref{three}).

Using (\ref{eqn:CKM}), every linear combination satisfying (\ref{psc}),
including (\ref{six}), (\ref{four}), and (\ref{three}), can be written as
\beq
\Sigma_f a_fA(f) = -\bl A'_P + \bl^{-1} A'_C~.
\eeq
One now forms the ratio of squared amplitudes, averaged over $B^0$ and
$\bar B^0$ and multiplied by $\bl^2$,
\bea\label{R}
{\cal R}^2 & \equiv &
\frac{\bl^2[|\Sigma_f a_fA(f)|^2 + |\Sigma_f a_f \bar A(f)|^2]}
{|A(B^0\to \eta' K^0)|^2 + |A(\bar B^0 \to \eta' \bar K^0)|^2}\nonumber\\
& = & \frac{|A'_C/A'_P|^2 + \bl^4 - 2\bl^2|A'_C/A'_P|\cos\delta\cos\gamma}
{1 + |A'_C/A'_P|^2 + 2|A'_C/A'_P|\cos\delta\cos\gamma}~,
\eea
where $\bar A(f)$ are decay amplitudes for a $\bar B^0$.
This expression may be inverted to become an expression for $|A'_C/A'_P|$:
\beq\label{A'c/A'p}
\frac{|A'_C|}{|A'_P|} = \frac{\sqrt{[(\bl^2 + {\cal R}^2)
\cos\delta\cos\gamma]^2 + (1-{\cal R}^2)({\cal R}^2 - \bl^4)} 
+ (\bl^2 + {\cal R}^2)\cos\delta\cos\gamma} {1-{\cal R}^2}~.
\eeq
Noting that $-1 \le \cos\delta\cos\gamma \le 1$,  one has
\beq
\frac{||A'_C/A'_P| - \bl^2|}{1 + |A'_C/A'_P|} \le {\cal R} \le
\frac{|A'_C/A'_P| + \bl^2}{1 - |A'_C/A'_P|}~,
\eeq
and
\beq\label{boundsA'c/A'p}
\frac{|{\cal R} - \bl^2|}{1 + {\cal R}}
\le |A'_C/A'_P| \le \frac{{\cal R} + \bl^2}{1 - {\cal R}}~.
\eeq

Upper bounds on ${\cal R}$ may be obtained from experiments using the general
algebraic inequality
\beq
|\Sigma_f a_fA(f)|^2 + |\Sigma_f a_f \bar A(f)|^2 \le
\left (\Sigma_f |a_f| \sqrt{|A(f)|^2 + |\bar A(f)|^2} \right )^2~.
\eeq
Denoting by $\bar {\cal B}_f$ branching ratios of $\Delta S=0$ decays,
averaged over $B^0$ and $\bar B^0$, and neglecting phase space differences
[in the spirit of assuming SU(3)] which can be included, one has
\beq\label{boundR}
{\cal R} \le \bl \Sigma_f |a_f| \sqrt{\frac{\bar {\cal B}_f}{\bar {\cal B}
(\eta'K^0)}}~.
\eeq
For a given set of coefficients $a_f$, nonzero branching ratio measurements
and upper limits on $\bar {\cal B}_f$ provide an upper bound on ${\cal R}$,
for which the right-hand-side of (\ref{boundsA'c/A'p}) gives an upper bound on
$|A'_C/A'_P|$.  As mentioned, the coefficients $a_f$ will be taken to have
values as in (\ref{six}), (\ref{four}), and (\ref{three}).

\begin{table}
\caption{Branching ratios in $10^{-6}$ and $90\%$ C.L. upper limits on
branching ratios}
\begin{center}
\begin{tabular}{c c c c c c c c c c} \hline
Mode  & $\eta' K^0$  & $\pi^0\pi^0$ & $K^+K^-$ &
$K^0\bar K^0$ & $\pi^0\eta$ & $\pi^0\eta'$ & 
$\eta\eta$ &
$\eta'\eta'$ & $\eta\eta'$ \\ \hline
$\bar {\cal B}$ &
$65.2 ^{+6.0}_{-5.9}$ & $1.9 \pm 0.5$ & $<0.6$ & $ < 1.5$ & $ < 2.5$ &
$ < 3.7$ & $< 2.8$ & $ < 10$ & $ < 4.6$ \\
CLEO  & & & & & $<2.9$ & $<5.7$ & $<18$ & $<47$ & $<27$ \\ \hline
\end{tabular}
\end{center}
\end{table}

Nonzero branching ratios, averaged over $B^0$ and $\bar B^0$, and $90\%$
confidence level upper limits on branching ratios~\cite{PDG} are listed in
Table I for $B^0 \to\eta' K^0$, and for the eight strangeness-conserving
processes occurring in Eqs.~(\ref{pieta}).  The last five measurements
involving $\eta$ and $\eta'$ were reported very recently by the BaBar
collaboration \cite{BaBar}. The second line in the Table lists earlier bounds
by the CLEO collaboration~\cite{CLEO}. The new bounds for the two processes
involving a $\pi^0$ and $\eta$ or $\eta'$ are only slightly stronger than the
earlier ones. However, bounds on the three processes involving pairs with
$\eta$ and $\eta'$ have improved considerably. 

We will now consider bounds on ${\cal R}$ obtained in the above three cases,
starting with a general SU(3) bound and continuing with bounds which neglect
small amplitudes:

\begin{enumerate}

\item Assuming exact SU(3) and applying (\ref{six}) we find, using the central
value for $\bar {\cal B}(\eta' K^0)$,
\beq\label{boundR1}
{\cal R} < 0.18~. 
\eeq
This strongest bound within pure SU(3) should be compared with a 
bound~\cite{GLNQ,CGR} $R < 0.36$ based on the earlier CLEO data,
and on an earlier upper limit, $\bar {\cal B}(\pi^0\pi^0)<5.7 \times 10^{-6}$
\cite{PDG}.   
In the exact SU(3) limit we also find that present data imply several 
almost degenerate minima for upper limits on ${\cal R}$, beside the 
point in parameter space describing the combination (\ref{six}). 

\item Applying (\ref{four}) one obtains
\beq\label{boundR2}
{\cal R} < 0.11~.
\eeq
This bound assumes a single SU(3) amplitude ($s$) in decays into two 
pseudoscalars belonging to an SU(3) singlet and an SU(3) octet. It should be
compared with ${\cal R} < 0.22$ obtained from the above-mentioned earlier data
using the same combination of amplitudes.

\item Neglecting $e$ and $pa$ and using (\ref{three}), which contains  
three processes, one finds
\beq\label{boundR3}
{\cal R} < 0.10~~. 
\eeq
This bound, which improves (\ref{boundR2}) only slightly, should be compared
with $R<0.18$ based on the earlier CLEO data.

\end{enumerate}
As mentioned above, the approximation involved in deriving (\ref{boundR3}), 
where SU(3) breaking and small amplitudes were neglected,  is comparable 
to that associated with (\ref{boundR1}) which only neglects SU(3) breaking 
effects.

When comparing upper limits on ${\cal R}$ implied by the recent BaBar
measurements with those obtained from the earlier CLEO measurements we observe
in all three approximations an improvement by a factor of about two.
In all three cases the present upper limit on $\bar {\cal B}(B^0 \to \eta
\eta')$ contributes the largest term.  Since (\ref{boundR1})--(\ref{boundR3})
were obtained by adding linearly experimental upper limits on $|a_f|\bar{\cal
B}_f^{1/2}$ at $90\%$ confidence level, and taking a central value for
$\bar{\cal B}(\eta'K^0)$ where the current error is less than $10\%$,
statistically these bounds involve a confidence level higher than $90\%$. 
However, their systematic uncertainties caused by SU(3) breaking and $e$ and
$pa$ corrections are expected to be at a level of 20--30$\%$.  

In order to study constraints in the ($S_{\eta' K}, C_{\eta' K}$) plane, we now
apply the upper bounds (\ref{boundR1}) and (\ref{boundR3}).  The exact
expressions (\ref{eqn:S})--(\ref{eqn:R}) imply correlated bounds on these two
quantities associated with fixed values of ${\cal R}$. We scan over $-\pi \le
\delta \le \pi$, taking a central value $\beta = 23.7^{\circ}$, values of
$\gamma$ satisfying $38^\circ \le \gamma \le 80^\circ$ \cite{CKMfitter}, and
values of $|A'_C/A'_P|$ in the range (\ref{boundsA'c/A'p}), where ${\cal R}$
satisfies the bound (\ref{boundR1}) or (\ref{boundR3}).  The ratio
$|A'_C/A'_P|$ does not saturate the bound (\ref{boundsA'c/A'p}) at the 
boundary of the allowed region, but is limited to smaller values, because
$\Delta S_{\eta'K}$ and $C_{\eta'K}$ in (\ref{eqn:S})--(\ref{SC}) are
approximately proportional to $\sin\gamma$ whereas $|A'_C/A'_P|$ in
(\ref{A'c/A'p}) increases with $\cos\gamma$.  The bounds on ($S_{\eta' K},
C_{\eta' K}$) are shown in Fig.~1.  Also shown are bounds based on the
earlier value ${\cal R} < 0.36$ mentioned above,
and two points corresponding to $(S_{\eta'K}, C_{\eta'K}) = (\sin 2\beta, 0)$
and $(0.75, -0.06)$ (see below).  

\section{Conclusion}

Our above discussion and Fig.~1 show that SU(3) bounds on the CP asymmetry of
$B^0\to \eta' K_S$ have improved considerably by incorporating the very recent
BaBar upper bounds and by neglecting small amplitudes, narrowing drastically
the region around $(S_{\eta'K}=\sin 2\beta, C_{\eta'K}=0)$ consistent with
the Standard Model.  A
critical and model-independent test of the Standard Model requires further
improvements, both in the asymmetry measurement and in branching ratio
measurements of $B^0$ decays into $K\bar K$ and into pairs involving
$\pi^0,~\eta$ and $\eta'$.  The theoretical bounds plotted in Fig.~1 are based
on branching ratio measurements of strangeness-conserving processes and on
flavor SU(3) considerations. These considerations, including neglecting small
$e$ and $pa$ amplitudes, may introduce theoretical errors in
the bounds on $\Delta S_{\eta'K}$ and $C_{\eta'K}$ of order 20 or 30 percent.

\begin{figure}[t]
\includegraphics[height=4.9in]{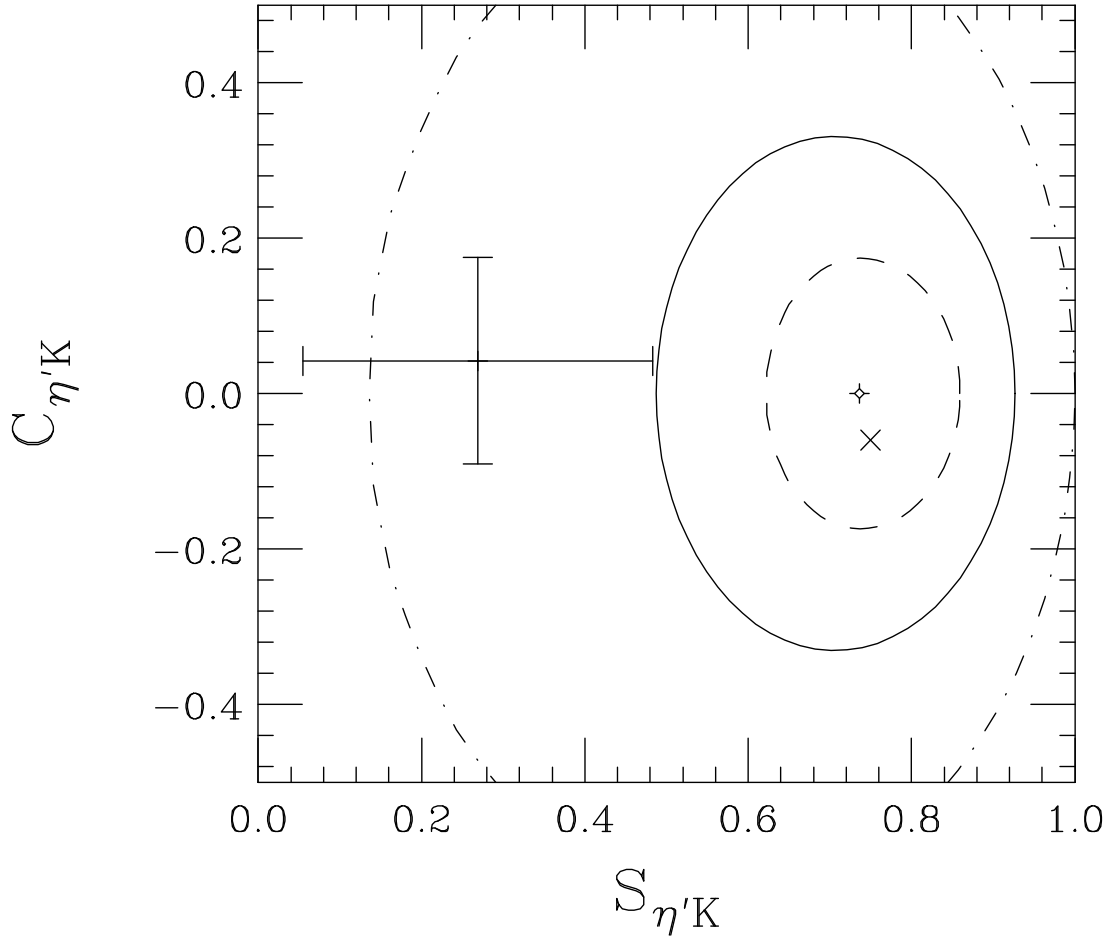}
\caption{Regions in the ($S_{\eta' K},~C_{\eta' K}$) plane satisfying
limits (25) on the ratio $|A'_C/A'_P|$ and bounds (\ref{boundR1}) (region
enclosed by the solid curve) or (\ref{boundR3}) (region enclosed by the dashed
curve).  The dot-dashed curve encloses the region satisfying an earlier bound
$R < 0.36$.  The plotted open point denotes $(S_{\eta' K},~C_{\eta' K}) = (\sin
2 \beta,~0)$, while the point labeled $\times$ denotes the central value of a
prediction in Ref.\ \cite{global}. 
\label{fig:maxetcpk}}
\end{figure}

These model-independent bounds are more conservative than other bounds which
involve further assumptions. A global SU(3) fit to all $B$ decays into pairs
of charmless pseudoscalar mesons (``Fit IV'' of Ref.\ \cite{global}) obtains
$|A'_C/A'_P| = 0.042^{+0.017}_{-0.006}$, corresponding to $0.034 <
|A'_C/A'_P| < 0.064$ at 90\% confidence level.  [In this case
the bounds (\ref{boundR1})--(\ref{boundR3}) are satisfied automatically.
A related fit (``III''), omitting one amplitude which improves a fit to
a single branching ratio, obtains $|A'_C/A'_P| = 0.040^{+0.011}_{-0.009}$.]
This would imply that the allowed area in the $(S_{\eta'K},~C_{\eta'K})$ plane
becomes smaller than the region corresponding to the bound $R < 0.10$.  In
Fig.\ 1 we show the value $(S_{\eta' K},~C_{\eta' K}) = (0.75^{+0.00}_{-0.01},
~-0.06^{+0.02}_ {-0.01})$ predicted in the favored Fit IV of Ref.\
\cite{global}.  [Fit III predicts $(S_{\eta' K},~C_{\eta' K}) = (0.74 \pm 0.01,
~-0.07 \pm 0.02)$.]  Smaller values of $|A'_C/A'_P|$ of order $0.01$ have been
calculated in~\cite{CGR,LS}, implying only tiny deviations at this level from
$S_{\eta'K}=\sin 2\beta,~C_{\eta'K}=0$.

\section*{Acknowledgments}

We thank Bill Ford, Fernando Palombo, Jim Smith, and Denis Suprun for helpful
comments.  J. L. R. thanks Maury Tigner for extending the hospitality of the
Laboratory for Elementary-Particle Physics at Cornell University and the
John Simon Guggenheim Memorial Foundation for partial support during this
investigation.  This work was supported in part by the United States Department
of Energy, High Energy Physics Division, through Grant No.\ DE-FG02-90ER40560. The work of J. Z. is supported in part by EU grant HPRN-CT-2002-00277 and  by the Ministry 
of Education, Science and Sport of the Republic of Slovenia.

\def \ajp#1#2#3{Am.\ J. Phys.\ {\bf#1}, #2 (#3)}
\def \apny#1#2#3{Ann.\ Phys.\ (N.Y.) {\bf#1}, #2 (#3)}
\def \app#1#2#3{Acta Phys.\ Polonica {\bf#1}, #2 (#3)}
\def \arnps#1#2#3{Ann.\ Rev.\ Nucl.\ Part.\ Sci.\ {\bf#1}, #2 (#3)}
\def \art{and references therein}
\def \cmts#1#2#3{Comments on Nucl.\ Part.\ Phys.\ {\bf#1}, #2 (#3)}
\def \cn{Collaboration}
\def \cp89{{\it CP Violation,} edited by C. Jarlskog (World Scientific,
Singapore, 1989)}
\def \econf#1#2#3{Electronic Conference Proceedings {\bf#1}, #2 (#3)}
\def \efi{Enrico Fermi Institute Report No.}
\def \epjc#1#2#3{Eur.\ Phys.\ J.\ C {\bf#1} (#3)  #2}
\def \ib{{\it ibid.}~}
\def \ibj#1#2#3{~{\bf#1}, #2 (#3)}
\def \ijmpa#1#2#3{Int.\ J.\ Mod.\ Phys.\ A {\bf#1}, #2 (#3)}
\def \ite{{\it et al.}}
\def \jhep#1#2#3{JHEP {\bf#1}, #2 (#3)}
\def \jpb#1#2#3{J.\ Phys.\ B {\bf#1}, #2 (#3)}
\def \mpla#1#2#3{Mod.\ Phys.\ Lett.\ A {\bf#1} (#3) #2}
\def \nat#1#2#3{Nature {\bf#1}, #2 (#3)}
\def \nc#1#2#3{Nuovo Cim.\ {\bf#1}, #2 (#3)}
\def \nima#1#2#3{Nucl.\ Instr.\ Meth.\ A {\bf#1}, #2 (#3)}
\def \npb#1#2#3{Nucl.\ Phys.\ B~{\bf#1} (#3) #2}
\def \npps#1#2#3{Nucl.\ Phys.\ Proc.\ Suppl.\ {\bf#1}, #2 (#3)}
\def \PDG{Particle Data Group, K. Hagiwara \ite, 
\prd{66}{010001}{2002}}
\def \pisma#1#2#3#4{Pis'ma Zh.\ Eksp.\ Teor.\ Fiz.\ {\bf#1}, 
#2 (#3) [JETP
Lett.\ {\bf#1}, #4 (#3)]}
\def \pl#1#2#3{Phys.\ Lett.\ {\bf#1}, #2 (#3)}
\def \pla#1#2#3{Phys.\ Lett.\ A {\bf#1}, #2 (#3)}
\def \plb#1#2#3{Phys.\ Lett.\ B {\bf#1} (#3) #2}
\def \prl#1#2#3{Phys.\ Rev.\ Lett.\ {\bf#1} (#3) #2}
\def \prd#1#2#3{Phys.\ Rev.\ D\ {\bf#1} (#3) #2}
\def \prp#1#2#3{Phys.\ Rep.\ {\bf#1}, #2 (#3)}
\def \ptp#1#2#3{Prog.\ Theor.\ Phys.\ {\bf#1} (#3) #2}
\def \rmp#1#2#3{Rev.\ Mod.\ Phys.\ {\bf#1}, #2 (#3)}
\def \yaf#1#2#3#4{Yad.\ Fiz.\ {\bf#1}, #2 (#3) [Sov.\ 
J.\ Nucl.\ Phys.\
{\bf #1}, #4 (#3)]}
\def \zhetf#1#2#3#4#5#6{Zh.\ Eksp.\ Teor.\ Fiz.\ {\bf #1}, 
#2 (#3) [Sov.\
Phys.\ - JETP {\bf #4}, #5 (#6)]}
\def \zpc#1#2#3{Zeit.\ Phys.\ C {\bf#1} (#3) #2}
\def \zpd#1#2#3{Zeit.\ Phys.\ D {\bf#1}, #2 (#3)}

\end{document}